\documentclass[preprint,showpacs,preprintnumbers,english]{revtex4}
\usepackage{amsmath}
\usepackage{graphicx}
\usepackage{dcolumn}
\usepackage{bm}

\input{tcilatex}
\begin{document}

\title{Spin convertance at magnetic interfaces}
\author{Steven S.-L. Zhang and Shufeng Zhang}
\affiliation{Department of Physics, University of Arizona, Tucson, AZ 85721}

\begin{abstract}
Exchange interaction between conduction electrons and magnetic
moments at magnetic interfaces leads to mutual conversion between
spin current and magnon current. We introduce a concept of spin
convertance which quantitatively measures magnon current induced by
spin accumulation and spin current created by magnon accumulation at
a magnetic interface. We predict several phenomena on charge and
spin drag across a magnetic insulator spacer for a few layered
structures.
\end{abstract}

\date{\today }
\maketitle

\section{Introduction}

In spintronics, spin current, which is conventionally defined as the
difference of electric currents of spin-up and spin-down conduction
electrons, plays a pivotal role in propagating spin information from
one place to another. Many spin dependent properties, such as giant
magnetoresistance \cite{Fert,Grunberg}, spin transfer torques \cite%
{Slonczewski,Berger} and spin Hall effect \cite{Hirsch,Zhang00}, are
directly related to spin current. Spin current has several unique
properties compared to charge current: 1) it is considered as a flow
of angular momentum while the conventional current is a flow of
charge, 2) the total spin current is not a conserved quantity even
in the steady state condition; it can be transferred and/or lost due
to spin-dependent scattering, and 3) spin current has both
transverse and longitudinal components whose decaying length scales
are quite different in a ferromagnetic medium. Recently, the concept
of spin current has been extended to spin wave current since spin
waves carry angular momenta as well \cite{Kajiwara10}. There are two
types of spin wave
currents. One is magnetostatic wave propagation \cite%
{Kajiwara10,Xiao12,Wu11,Nowak11,Wang11} for which the classical
magnetization is temporal and spatially dependent. An example is a
moving domain wall driven by a magnetic field or by an electric
current. Although such magnetostatic spin waves may carry angular
momentum, they are not quasi-particles in that there are no particle
numbers associated with these waves. The other spin wave current is
a true quasi-particle current known as magnon current. A magnon is a
quantum object (particle) that represents low excitation state of
ferromagnets. In equilibrium, the number of magnons
$N_{\mathbf{q}}$ can be cast into a simple Boson distribution $N_{\mathbf{q}%
}^{0} = [e^{E_{\mathbf{q}}/k_BT}-1]^{-1}$ where $E_{\mathbf{q}}$ is
the magnon energy. Similar to the electron spin, each magnon carries
an angular momentum $-\hbar$. In thermal equilibrium, there is no
magnon current since there are an equal number of magnons moving in
all directions.

In our earlier paper \cite{Zhang12}, we showed that the
non-equilibrium magnon accumulation and magnon current can be
treated semiclassically, similar to the spin transport properties of
conduction electrons. We found that the non-equilibrium electron
spin current in metal can convert into a magnon current of a
magnetic insulator through the interfacial exchange interaction. The
magnon current then subsequently diffuses inside the magnetic
insulator. The magnon diffusion process may be described by the
diffusion equation. Among other things, we predicted that an
electric current applied in one metallic layer can induce an
electric current in another metallic layer separated by a magnetic
insulator via magnon mediated angular momentum transfer. In this
paper, we extend our theory to include a general boundary condition
for the spin convertibility at metal$\mid$magnetic-insulator
interfaces and then, we calculate the electric drag in a few
realizations. The paper is organized as follows. In Sec.~II, we
summarize the general boundary conditions at the interfaces between
metals and magnetic insulators. In particular, we introduce a
quantity, named as spin convertance, which quantitatively
characterizes conversion effectiveness among spin/magnon
accumulation and magnon/spin current at a magnetic interface. In
Sec.~III, we calculate the spin convertance by using the microscopic
s-d exchange interaction. In Sec.~IV, we present the general
solutions for several layered structures with a magnetic insulator
layer (MIL) and discuss some limiting cases. Finally, we summarize
our results.

\section{Summary of boundary conditions}

We consider a simple bilayer consisting of a metallic layer in
contact with a MIL. The angular momentum in a metal is carried by
conduction electrons while in a MIL, it is carried by magnons. In
the semiclassical approximation, spin transport properties can be
described by the Boltzmann distributions of electrons and magnons
\cite{Zhang12}. The boundary conditions are to link the
non-equilibrium electron distribution function of the metal to the
magnon distribution function of the MIL. Within the model of the s-d
exchange interaction (see Sec.~III), the total angular momentum is
conserved and thus for an ideal interface the first boundary
condition would be
\begin{equation}
j_{t} (0^-) = j_{t}(0^+)
\end{equation}
where $j_{t}$ is the total angular momentum current, and we assign the
interface at $x=0$. If we consider the left layer as a non-magnetic metal ($%
x<0$), the angular momentum is carried by conduction electrons only
and thus $j_t(0^-) = j_s(0^-)$ where $j_s$ denotes the conventional
spin current density. For the MIL on the right, the angular momentum
is carried by magnons only and thus $j_t(0^+) =j_m(0^+)$ where $j_m$
corresponds to the magnon current density. Therefore, we may rewrite
Eq.~(1) as
\begin{equation}
j_{s} (0^-) = j_{m}(0^+)
\end{equation}
Note that for a magnetic metal, both spin and magnon current contribute to
the total angular momentum current.

The second boundary condition is the relation among the electron
spin accumulation $\delta m_s (0^-)$, the magnon accumulation
$\delta m_m (0^+)$, and the total angular momentum current $j_{t}
(0)$,
\begin{equation}
G_{em}\delta m_s (0^-) - G_{me} \delta m_m (0^+) = j_{t} (0)
\end{equation}
where the two coefficients $G_{em}$ and $G_{me}$ will be calculated
within the s-d model in the next section. The physics of this
boundary condition is rather transparent: the first term represents
the generation of the magnon current in the presence of electron
spin accumulation and the second term describes the spin current
produced by magnon accumulation. The combination of these two
processes at the interface yields the total interface spin current.
We immediately note that Eq.~(3) is analogous to the case of spin
current between two metallic layers in which the boundary condition
is $G_{\sigma} \mu_{\sigma} (0^+) - G_{\sigma} \mu_{\sigma} (0^-)
=j_{\sigma} (0)$ where $\mu_{\sigma}$ denotes the spin dependent
chemical potential ($\sigma=\pm1$ or $\uparrow (\downarrow)$
corresponds to spin-up (down)) which is proportional to spin
accumulation, and $G_{\sigma}$ characterizes the interfacial spin
conductance \cite{Valet93}. With this analogy, we may identify the
coefficients $G_{em}$ ($G_{me}$) as the interface conductance for
the conversion of the spin (magnon) accumulation to the magnon
(spin) current; we simply call $G_{em}$ and $G_{me}$ \emph{spin
convertance} for convenience hereafter.

We point out that the boundary condition, Eq.~(3), is different from
what we proposed in the earlier paper \cite{Zhang12} where we
related the spin and magnon accumulation via a local magnetic
susceptibility. Clearly, such approximation corresponds to an ideal
case in which the interface spin resistance is zero or the spin
convertance is infinite. In the next section, we shall calculate
these spin convertances and show that they are
in fact finite and thus the magnon mediated electric drag effect predicted in Ref.~%
\cite{Zhang12} was overestimated by one order of magnitude.

\section{Microscopic calculation of spin convertance}

We start with the s-d exchange coupling at a metal$\mid$MIL
interface,
\begin{equation}
\hat{H}_{sd}=-J_{sd}\sqrt{\frac{S}{2N_{s}}}\underset{\mathbf{k},\mathbf{q},%
\mathbf{k}\prime }{\sum }(a_{\mathbf{q}}^{\dag }c_{\mathbf{k}\uparrow
}^{\dag }c_{\mathbf{k}\prime \downarrow }+a_{\mathbf{q}}c_{\mathbf{k}\prime
\downarrow }^{\dag }c_{\mathbf{k}\uparrow })\delta _{\mathbf{k}\prime =%
\mathbf{q+k}}
\end{equation}%
where $c_{\mathbf{k}\uparrow }^{\dag }$ ($c_{\mathbf{k}\uparrow }$)
and $c_{\mathbf{k}\downarrow }^{\dag }$ ($c_{\mathbf{k}\downarrow
}$) are the creation (annihilation) operators for spin-up and spin
spin-down electrons respectively, $a_{\mathbf{q}}^{\dag}$
($a_{\mathbf{q}}$) is the creation (annihilation) operator for magnons, $S$ is the spin per atom of the MIL, and $%
N_s$ is the number of atomic spins of the MIL at the interface. The
exchange coupling strength $J_{sd}$ is given by the exchange
integral with the overlapped wavefunctions of the conduction
electrons and the magnetic ions. Since we do not know the detailed
orbitals for the interface states, the magnitude of $J_{sd}$ at
interface is less known compared to that in bulk materials and we
will treat it as a parameter.

The above exchange interaction gives rise to angular momentum
transfer between the electron spins at the metallic side and the
magnons at the MIL side. In equilibrium, the net spin current across
the interface is zero. At non-equilibrium when there is a spin
accumulation at $x=0^-$ or a magnon accumulation at $x=0^+$, a net
magnon/spin current may be present across the interface. In Fig.~1,
we illustrate two angular momentum transfer processes. The total
angular momentum current across the interface should be caused by
both processes. We shall calculate them separately below.

\begin{figure}[tbp]
\includegraphics[width=8cm]{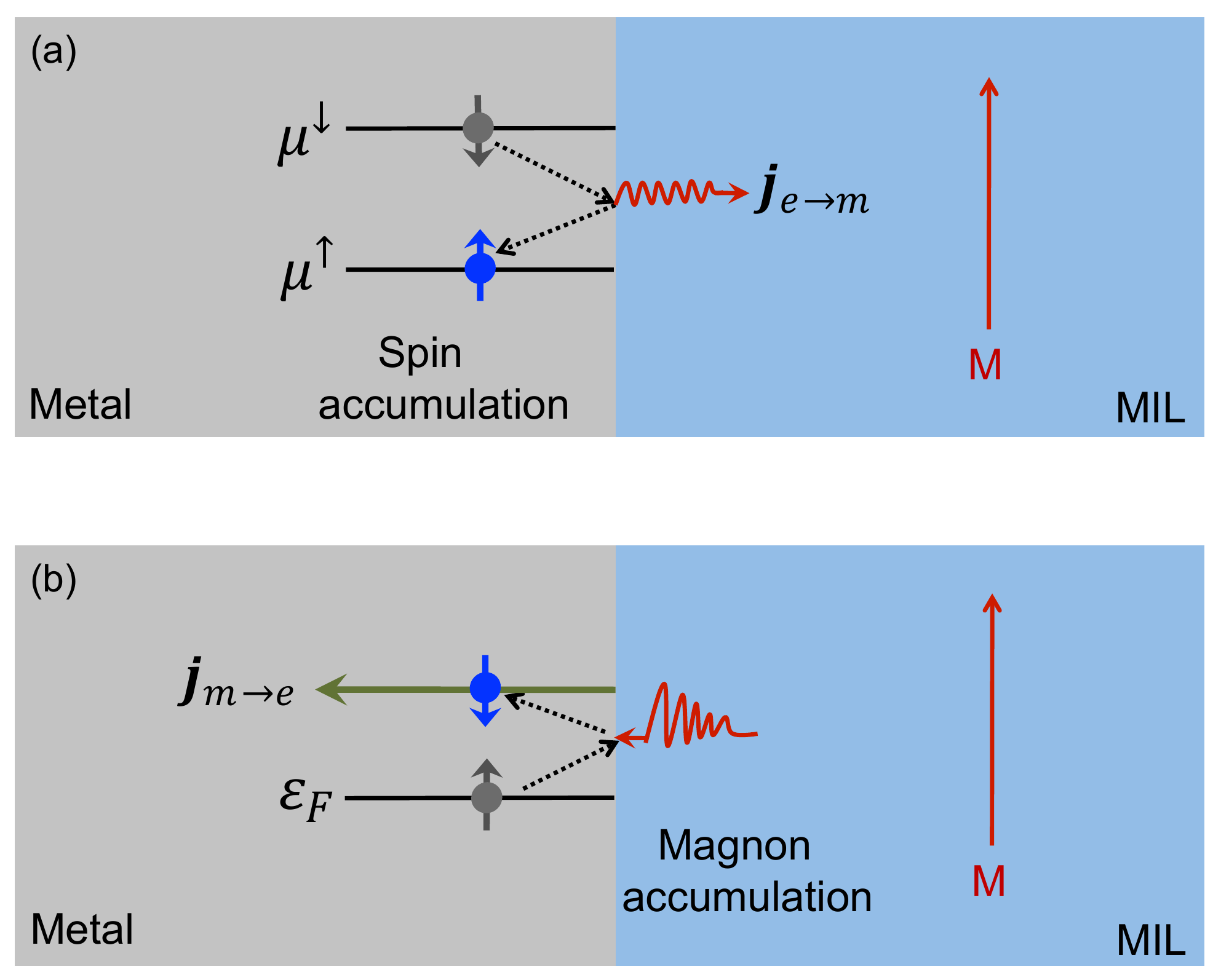}
\caption{Spin angular momentum transfer at a metal$\mid$MIL
interface. The upper (lower) panel describes magnon current
$j_{e\rightarrow m}$ (spin
current $j_{m\rightarrow e}$) generated by spin accumulation $%
\protect\delta m_s$ (magnon accumulation $\protect\delta m_m$) at the
interface.}
\end{figure}

Magnon current generated by spin accumulation at the interface is
defined as
\begin{equation}
j_{e\rightarrow m}\equiv \left\langle \frac{\mu _{B}}{i\hbar A_{I}}\left[
\sum_{\mathbf{k}}(c_{\mathbf{k}\uparrow }^{\dag}c_{\mathbf{k}\uparrow }-c_{%
\mathbf{k}\downarrow }^{\dag}c_{\mathbf{k}\downarrow
}),\hat{H}_{sd}\right] \right\rangle
\end{equation}%
where $<>$ refers to the thermal averaging over all states and
$A_{I}$ is the area of the interface. By explicitly working out the
above commutator and by using the Fermi-golden rule, we have
\begin{equation}
j_{e\rightarrow m}=\frac{2\pi \mu _{B}S}{A_{I}\hbar }\left( \frac{J_{sd}}{%
N_{s}}\right) ^{2}\underset{\mathbf{k},\mathbf{q},\mathbf{k}\prime }{\sum }%
\left[ (N_{\mathbf{q}}+1)(1-f_{\mathbf{k}\uparrow })f_{\mathbf{k}\prime
\downarrow }-N_{\mathbf{q}}(1-f_{\mathbf{k}\prime \downarrow })f_{\mathbf{k}%
\uparrow }\right] \delta (\varepsilon _{\mathbf{k}}+E_{q}-\varepsilon _{%
\mathbf{k}\prime })
\end{equation}%
where $N_{\mathbf{q}}$ and $f_{\mathbf{k}\sigma }$ are the magnon
and electron distribution functions respectively. We have considered
a rough interface such that there is no correlation between the
electron and magnon momenta
for the magnon emission/absorption processes (i.e., we do not impose $\mathbf{k}%
^{\prime }-\mathbf{k}=\mathbf{q}$). We first consider the process in
the upper panel of Fig.~1, i.e., magnon current due to electron spin
accumulation. Accordingly, we take the equilibrium distribution
function for magnons,
i.e., $N_{\mathbf{q}}=N_{\mathbf{q}}^{0}=[\exp (E_{\mathbf{q}%
}/k_{B}T)-1]^{-1}$ where the spin-wave energy is $E_{\mathbf{q}%
}=A\mathbf{q}^{2}+\Delta _{g}$, the exchange stiffness is associated
with the
Curie temperature via $%
A=3k_{B}T_{c}a_{0I}^{2}/\pi ^{2}(S+1)$ \cite{Zhang&Parkin97}, and
$\Delta _{g}$ is the spin wave gap due to magnetic anisotropy. The
electron distribution function can be conveniently separated into
equilibrium and non-equilibrium parts,
\begin{equation}
f_{\mathbf{k}\sigma }=f_{\mathbf{k}}^{0}+\frac{\partial f_{\mathbf{k}}^{0}}{%
\partial \varepsilon _{\mathbf{k}}}[-\delta \mu_{\sigma }(x)+g_{\sigma }(%
\mathbf{k},x)]
\end{equation}%
where $f_{\mathbf{k}}^{0}$ is the Fermi distribution function, $\delta \mu
_{\sigma }(x)$ is the local variation of the chemical potential and $%
g_{\sigma }(%
\mathbf{k},x)$ is the anisotropic part of the non-equilibrium
distribution function ($\int d^{3}\mathbf{k}g_{\sigma
}(\mathbf{k},x)=0$). By placing the
above \emph{equilibrium} magnon distribution function and \emph{%
non-equilibrium} electron distribution function into Eq.~(6), we arrive at
\begin{equation}
j_{e\rightarrow m}=G_{em}\delta m_{s}(0^{-})
\end{equation}%
where we have defined $\delta m_{s}=\mu _{B}g_{e}({\epsilon
_{F}})(\delta \mu_{\uparrow }-\delta \mu_{\downarrow })$ as the spin
accumulation with $g_{e}({\varepsilon_{F}})$ being the interface
electron density of states at Fermi level. The spin convertance can
be formulated by
\begin{equation}
G_{em} =\frac{\pi S}{2\hbar k_B T }J_{sd}^{2}g_{e}(\varepsilon
_{F})a_{0M}^{2}a_{0I}^{5}\int\limits_{\Delta _{g}}^{E_{max}}dE_{\mathbf{q}}g_{m}(E_{%
\mathbf{q}})E_{\mathbf{q}}\mathrm{csch}^2\left(
\frac{E_{\mathbf{q}}}{2k_{B}T}\right)  % \notag
%\\
%&\approx &\frac{\sqrt{3}\pi ^{2}S(S+1)^{\frac{3}{2}}}{6\hbar }%
%J_{sd}^{2}g_{e}(\varepsilon _{F})a_{0M}^{4}\left( \frac{T}{T_{c}}\right) ^{%
%\frac{3}{2}}.
\end{equation}%
where $a_{0M}$ and $a_{0I}$ are the lattice constants of the metal
layer and the MIL respectively, $g_{m}(E_{\mathbf{q}})$ is the
interface magnon density of states, and $E_{max}$ ($\simeq
3k_BT_c/(S+1)$) is the maximum magnon energy. If a parabolic magnon
dispersion is assumed, then the dominant temperature dependence of
$G_{em}$ is $(T/T_c)^{3/2}$. The above result has already been
obtained in \cite{Takahashi10, Tserkovnyak12}.

The spin current induced by magnon accumulation at the
metal$\mid$MIL interface can be similarly calculated. We define this
interface spin current as
\begin{equation}
j_{m\rightarrow e}\equiv \left\langle \frac{2\mu _{B}}{i\hbar
A_{I}}\left[
\sum_{\mathbf{q}}a_{\mathbf{q}}^{\dag}a_{\mathbf{q}},\hat{H_{sd}}\right]
\right\rangle .
\end{equation}%
After working out the ensured commutator, we find the spin current
has exactly the same expression as Eq.~(6); this is not surprising
because the s-d interaction conserves the total angular momenta. To
evaluate the spin current induced by magnon accumulation (see the
process displayed in the lower panel of Fig.~1), we replace the
electron distribution by the equilibrium value, $f_{\mathbf{k}\sigma
}=f_{\mathbf{k}}^{0}$, and
separate the magnon density into equilibrium and non-equilibrium ingredients $N_{\mathbf{%
q}}=N_{\mathbf{q}}^{0}+\delta N_{\mathbf{q}}$. We then find from
Eq.~(10),
\begin{equation}
j_{m\rightarrow e}=G_{me}\delta m_{m}(0^{+})
\end{equation}%
where $\delta m_{m}\equiv (2\mu
_{B})\int\limits_{{}}^{{}}dE_{\mathbf{q}}g_{m}(E_{\mathbf{q}})\delta
N_{\mathbf{q}}$ is defined as the magnon accumulation. The spin
convertance can be expressed as
\begin{equation}
G_{me}=\frac{\pi S}{\hbar }J_{sd}^{2}g_{e}^{2}(\varepsilon
_{F})a_{0M}^{2}a_{0I}^{5}\bar{E}_{m}
\end{equation}%
with
\begin{equation}
\bar{E}_{m}=\frac{\int\limits_{\Delta _{g}}^{E_{max}}dE_{\mathbf{q}}g_{m}(E_{\mathbf{q}})E_{\mathbf{q}}%
N_{\mathbf{q}}^{0}}{\int\limits_{\Delta
_{g}}^{E_{max}}dE_{\mathbf{q}}g_{m}(E_{\mathbf{q}})N_{\mathbf{q}}^{0}}
\end{equation}%
where we have replaced the non-equilibrium magnon energy by the
average magnon energy $\bar{E}_{m}$ by assuming a near equilibrium
magnon distribution. A rough estimation for simple parabolic bands
of both magnons and electrons gives $G_{me}\sim (\pi S
a_{0I}^{5}/\hbar a_{0M})J_{sd}^{2}g_{e}(\varepsilon
_{F})\left(\frac{T}{T_{F}}\right)$ where $T_{F}$ is the Fermi
temperature of the metal layer.

By combining Eq.~(8) and (11), we attain Eq.~(3) with the spin
convertances $G_{em}$ and $G_{me}$ given by Eqs.~(9) and (12).

\section{Role of spin convertance in electrical drag}

To experimentally realize the conversion between spin current and
magnon current and to quantify the spin convertance, one needs to
create a non-equilibrium condition such that a spin current or a
magnon current can be generated, manipulated, and more critically,
detected. The non-equilibrium states may be created in several ways.
In this section, we study both electrical injection into a metal
layer and thermal gradient across a MIL. In Fig.~2, we show three
hypothetical devices to explicitly demonstrate the magnon-mediated
electrical drag.

\begin{figure}[tbp]
\includegraphics[width=16cm]{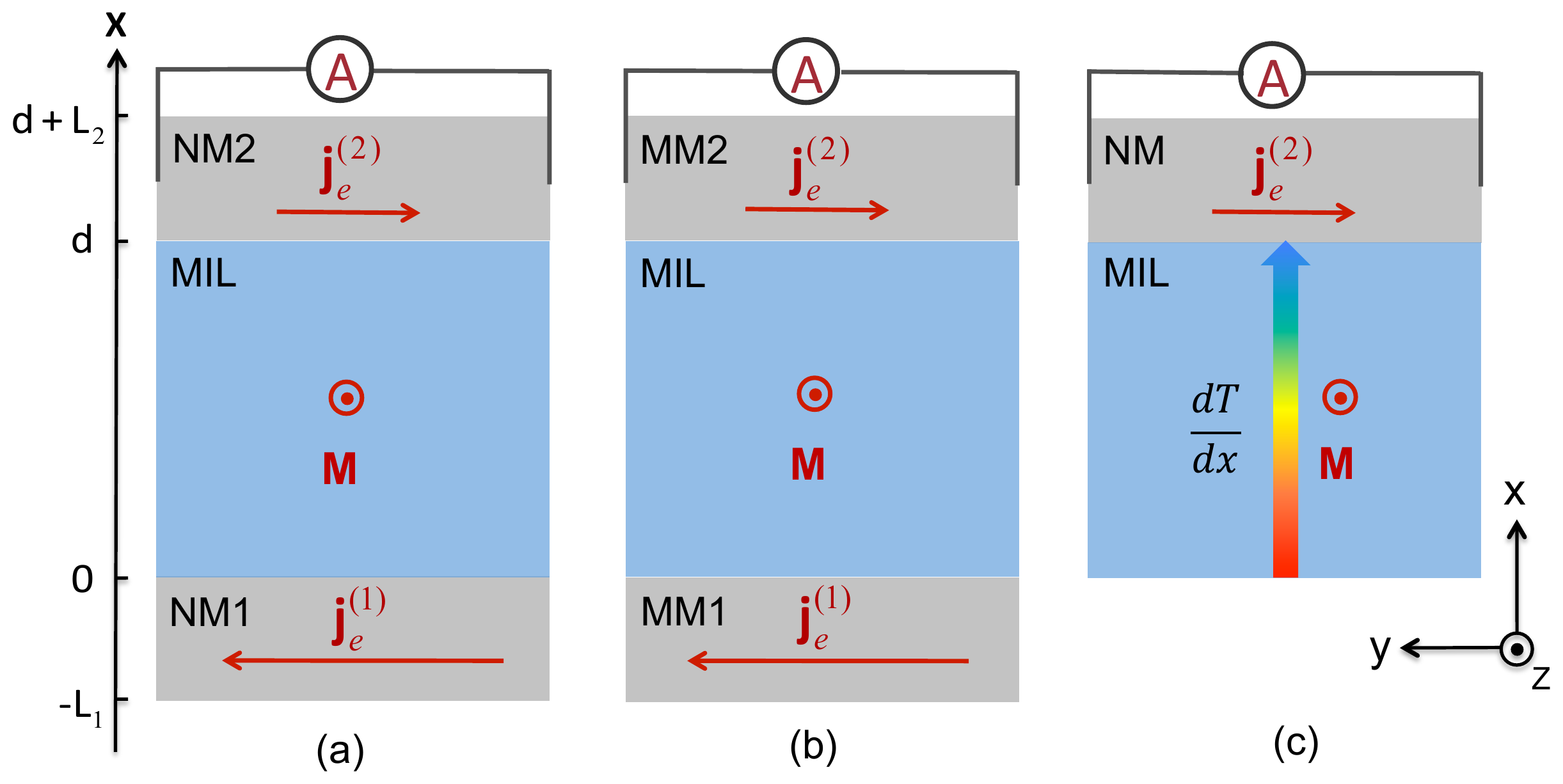}
\caption{Schematics of three hypothetical devices: (a)
NM$\mid$MIL$\mid$NM trilayers, (b) MM$\mid$MIL$\mid$MM trilayers,
and (c) NM$\mid$MIL bilayers. In (a) and (b), a spin current is
generated by an injected electric current via spin Hall effect,
while in (c), a magnon current is induced by applying a thermal
gradient. In all three cases, the magnetization directions of the
MIL and the magnetic metal (MM) layers are oriented in $+z$.}
\end{figure}

\subsection{NM$\mid$MIL$\mid$NM trilayers}

In Fig.~2(a), a magnetic insulator layer (MIL) is sandwiched between
two non-magnetic metal (NM) layers. By applying an in-plane
electrical current in the NM1 layer, a spin current flowing
perpendicular to the layers would be generated due to the spin Hall
effect. In this geometry, a partial spin current would flow into the
MIL via transfer of spin current to magnon current. If the magnon
diffusion length is larger than the thickness of the MIL, the magnon
current would reach the other side of the MIL and subsequently,
converts back to spin current in the NM2 layer. Finally, an electric
current parallel
to the layer is generated owing to the inverse spin Hall effect \cite%
{Saitoh06}. Such an electric drag phenomenon, namely, an electric
current in one NM layer induces an electric current in the other
when the two NM layers are separated by a MIL, would be a proof of
the magnon/spin current conversion at magnetic interfaces. Although
we have already calculated the drag coefficient in
Ref.~\cite{Zhang12}, we find the improved boundary conditions
presented in this paper quantitatively modify the earlier result.

Referring to the coordinate system in Fig.~2, one can establish the
relation of the spin accumulation, spin current, magnon
accumulation, and magnon current in each layer. For the NM1 layer
with an applied in-plane current density $j^{(1)}_{e}$, we have
\begin{equation}
\delta m_{s}(x)=A_{1}\exp (x/\lambda _{sf})
\end{equation}%
where $\lambda _{sf}$ is the spin diffusion length, $A_{1}$ is a
constant to be determined via boundary conditions. We have taken the
thickness of the layer much larger than the spin diffusion length
such that the term proportional to $\exp(-x/\lambda_{sf})$ has been
dropped. The spin current flowing perpendicular to the plane of the
layers is given by
\begin{equation}
j^{(1)}_{s}(x)= -\gamma_{sh}j^{(1)}_{e}-D_{s}\frac{\partial \delta m_{s}}{%
\partial x}
\end{equation}%
where the first term represents the spin Hall effect: an electric current $%
j_{e}^{(1)}$ in the $y$-direction generates a transverse spin
current proportional to the spin Hall angle $\gamma_{sh}$ which is
defined as the ratio of the spin Hall conductivity to the electric
conductivity. Note that we have adopted $e=\mu_B=1$ for notation
convenience so that the electrical current and the spin current
would have the same unit. The second term corresponds to the spin
diffusion where $D_{s}$ is the spin diffusion coefficient which may
be related to the
conductivities $c_{\uparrow}=c_{\downarrow }$ by the Einstein relation: $%
c_{\uparrow(\downarrow)}=e^{2}g_{e}(\varepsilon _{F})D_{s}$. For the
MIL layer, we have
\begin{equation}
\delta m_{m}(x)=A_{2}\exp (x/l_{m})+A_{3}\exp (-x/l_{m})
\end{equation}%
and
\begin{equation}
j_{m}(x)=-D_{m}\frac{\partial \delta m_{m}}{\partial x}
\end{equation}%
where $A_{2}$ and $A_{3}$ are integral constants from the magnon diffusion
equation, $l_{m}$ is the magnon diffusion length, $D_{m}$ is the magnon
diffusion constant associated with the magnon diffusion length by $%
l_{m}=\sqrt{D_{m}\tau _{th}}$ with $\tau _{th}$ being the
magnon-nonconserving relaxation time \cite{Zhang12}. For the NM2
layer, we have
\begin{equation}
\delta m_{s}(x)=A_{4}\exp (x/\lambda _{sf})+A_{5}\exp (-x/\lambda
_{sf})
\end{equation}%
and
\begin{equation}
j^{(2)}_{s}(x)=-D_{s}\frac{\partial \delta m_{s}}{\partial x}
\end{equation}%
where $A_4$ and $A_5$ are two integration constants. The four boundary conditions of Eqs.~(2) and (3) at the two interfaces $x=0$ and $x=d$%
, along with the outer-boundary condition at $x=d+L_{2}$ where $%
j^{(2)}_{s}(x=d+L_{2})=0$, determine the five constants $A_{i}$
($i=1-5$). After a straighforward algebra, we find the spin current
density $j^{(2)}_s (x)$ which in turn converts to an in-plane charge
current in the NM2 layer via the inverse spin Hall effect, i.e.,
$j_e^{(2)}(x) = \gamma_{sh} j^{(2)}_s(x)$. Explicitly,
\begin{equation}
j_{e}^{(2)}(x)=\frac{-a b \sinh \left[ \frac{d+L_{2}-x}{\lambda_{sf}} \right]
\mathrm{csch} \left( \frac{L_{2}}{\lambda _{sf}}\right)
\gamma_{sh}^{2}j_{e}^{(1)}}{ \left[ b_1 +b_2 \coth \left( \frac{L_{2}}{%
\lambda _{sf}}\right) \right] \sinh \left( \frac{d}{l_{m}}\right) +\left[
b_3 + ab\coth \left( \frac{L_{2}}{\lambda _{sf}}\right) \right] \cosh \left(
\frac{d}{l_{m}}\right) }
\end{equation}
where we have introduced the dimensionless constants $a\equiv
\lambda_{sf} G_{em}/D_s $, $b\equiv {l_m G_{me}}/D_m $ $b_1 =
1+a+b^2$, $b_2 =a+a^2$, and $b_3 = (2+a)b $. We may define an
average electric current density by averaging over the thickness of
the NM2 layer, $\bar{j}_{e}^{(2)}=(1/L_{2})\int dxj_{e}^{(2)}(x)$.
Then the ratio of the averaged current density to the injected
current density, i.e., $\eta \equiv \left\vert
\bar{j}_{e}^{(2)}/j_{e}^{(1)}\right\vert $, can be obtained as

\begin{equation}
\eta = \frac{\lambda _{sf} \gamma^2_{sh}}{L_2} \frac{ab \tanh \left(\frac{%
L_{2}}{2 \lambda_{sf}}\right)}{ \left[ b_1 +b_2 \coth \left( \frac{L_{2}}{%
\lambda _{sf}}\right) \right] \sinh \left( \frac{d}{l_{m}}\right) +\left[
b_3 +ab \coth \left( \frac{L_{2}}{\lambda _{sf}}\right) \right] \cosh \left(
\frac{d}{l_{m}}\right) }
\end{equation}
The electrical drag coefficient $\eta$ may be readily estimated. In
the case of $d \ll l_m$, $\eta$ becomes independent of $G_{me}$, but
increases with $G_{em}$; this is understandable since in this case
the magnon current does not decay and thus the magnon accumulation
is unimportant, $\eta$ depends predominantly on the efficiency of
the magnon current generation by spin accumulation which is measured
by $G_{em}$. A quick numerical check also indicates that $G_{em}$ is
usually larger than $G_{me}$. We consider a trilayer of
Ta$\mid$YIG$\mid$Ta whose material parameters at room temperature
($T=300 K$) are taken
as follows: for the Ta layers \cite{Buhrman12}, the conductivity $%
c_{Ta}= \left(190 \mu \Omega \cdot cm\right) ^{-1}$, the spin
diffusion length $\lambda_{sf}=5$ $nm$ and the spin Hall angle
$\gamma_{sh}=0.15$, the
lattice constant $a_{0M}=3.3{\mathring{A}}$, and the Fermi energy $\varepsilon_F=5$ $%
eV$; for the YIG layer \cite{Callen65}, the Curie temperature $%
T_{c}=550$ $K$, the lattice constant $a_{0I}=12.376{\mathring{A}}$,
the spin wave
gap $\Delta_g=10^{-6}$ $eV$, and the magnon relaxation time $\tau _{th}=10^{-6}s$%
. In Fig.~3, we show $\eta $ as a function of the thicknesses of the
NM2 layer for several different MIL thicknesses. Fig.~4 shows $\eta
$ as a function of the interface exchange coupling $J_{sd} $ with
several different magnon diffusion lengths.

Finally, we discuss the sign of the drag current. \emph{The induced
electric current always flows in the opposite direction of the
injected electric current for any magnetization direction of the MIL
}. To see this, we first recall the spin Hall and inverse spin Hall
effect in a single layer: an electric current induces a
perpendicular spin current (spin Hall) which in turn produces an
electric current (inverse spin Hall). The physical principle is that
the combined actions of the spin Hall and the inverse spin Hall are
to reduce the original driving current. Now consider the trilayer
system. Since the spin current injected into the NM2 layer remains
parallel to the spin current in the NM1 layer, the electric drag
current in the NM2 layer must be antiparallel to the applied
electric current in the NM1 layer.

\begin{figure}[tbp]
\includegraphics[width=8cm]{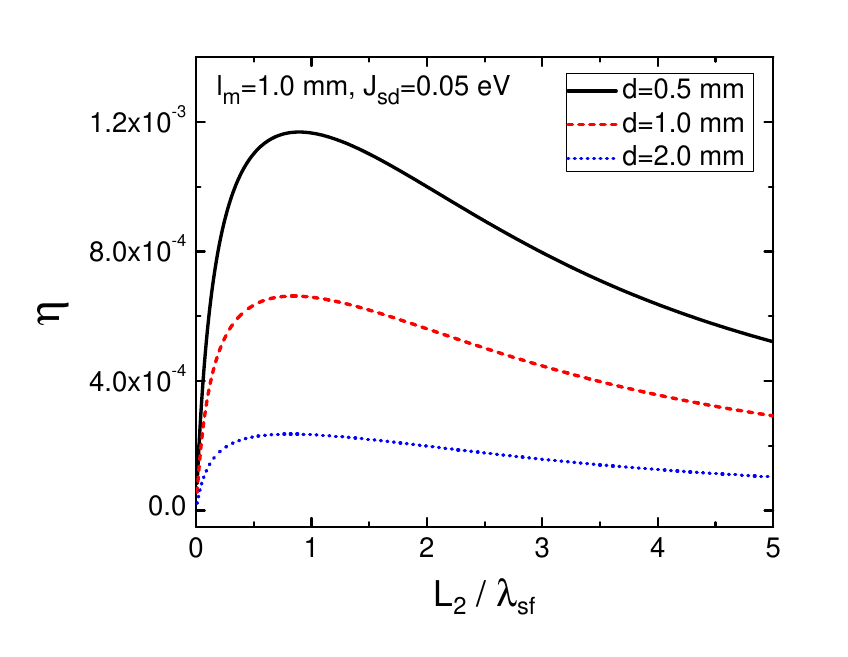}
\caption{Electrical drag coefficient as a function of the NM2 (Ta)
layer thickness for three different thicknesses of the MIL (YIG).
See the main text for the parameters used in the figure.}
\end{figure}

\begin{figure}[tbp]
\includegraphics[width=8cm]{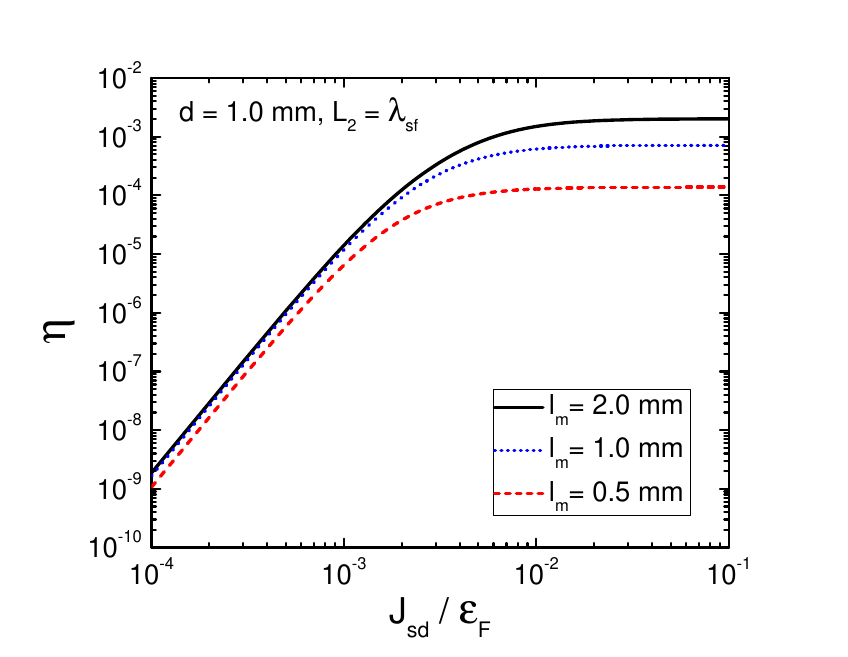}
\caption{Electrical drag coefficient as a function of $J_{sd}/\protect%
\varepsilon_F$ for three different magnon diffusion lengths of the
MIL (YIG). See the main text for the parameters used in the figure.}
\end{figure}

\subsection{MM$\mid$MIL$\mid$MM trilayers}

Next, we consider a trilayer structure where the two metallic layers
are magnetic, as shown in Fig.~2(b). Since the direct contact
between the magnetic metal (MM) layer and the MIL would make it
difficult to rotate the magnetization of each layer independently,
one may insert a thin non-magnetic layer at the interface to break
direct magnetic coupling. When an in-plane current is applied to the
MM1 layer, an anomalous Hall current perpendicular to the layers is
generated if the magnetization is oriented in the $z$-axis. Although
the physics of anomalous Hall and spin Hall effects are the same,
the anomalous Hall current has both spin and charge currents. The
charge current, however, is unable to penetrate the MIL; this leads
to a charge accumulation at the interface so that the net charge
current is exactly zero in the steady state. The spin current, on
the other hand, is able to propagate into the MIL via the conversion
to the magnon current, as discussed in the previous section. To gain
a quantitative understanding, we carry out the following
calculation.

The $x$-components of spin and charge currents of the MM1 layer can
be expressed as
\begin{equation}
j_{s}^{x (1)}=- p D_s \frac{\partial \delta n_{0}^{(1)}}{\partial x}-D_{s}\frac{%
\partial \delta m_{s}^{(1)}}{\partial x}- \gamma_{ah}j_e^{(1)}
\end{equation}
and
\begin{equation}
j_{e}^{x(1)}=-D_s \frac{\partial \delta n_{0}^{(1)}}{\partial x}- p D_s\frac{%
\partial \delta m_{s}^{(1)}}{\partial x}- p \gamma_{ah}j_e^{(1)}
\end{equation}
where
$p=(c_{\uparrow}-c_{\downarrow})/(c_{\uparrow}+c_{\downarrow})$ is
the spin polarization of the conductivity, $\gamma_{ah}$ is the
anomalous Hall angle defined as the ratio of the Hall conductivity
to the longitudinal conductivity, $\delta n_0$ is the charge
accumulation and $j_e^{(1)}$ is the current density applied in the
$y$-direction as before. We have assumed a spin-independent spin
diffusion coefficient $D_s$. Since $j_{e}^{x(1)}=0$, we may
eliminate the charge accumulation term from Eq.~(22) and get
\begin{equation}
j_{s}^{x(1)}=-(1-p^{2})D_{s}\frac{\partial \delta
m_{s}^{(1)}}{\partial x}- (1-p^2)\gamma_{ah}j_e^{(1)}
\end{equation}
For the MIL, Eqs.~(16) and (17) remain valid, while for the MM2
layer, we similarly have
\begin{equation}
j_{s}^{x(2)}=-(1-p^{2})D_{s}\frac{\partial m_{s}^{(2)}}{\partial x}
\end{equation}
By comparing Eqs.~(24) and (25) with Eqs.~(15) and (19), one should
realize that the induced electric current $j_{e}^{(2)}$ in the MM2
layer can be simply obtained by replacing $D_s$ by $(1-p^2)D_s$ and
$\gamma_{sh}$ by $(1-p^2)\gamma_{ah}$ in Eq.~(20). Consequently, the
electrical drag current is reduced by a fact of $(1-p^2)^2$ for the
MM$\mid$MIL$\mid$MM trilayer if one approximates $\gamma_{sh}
\approx \gamma_{ah}$. This might be counter-intuitive at first
glance, since one would expect the magnetic metals to provide more
spin signals. However, if we realize the interplay between the
charge and spin currents, one can readily explain the above
conclusion: consider the extreme case of $p=1$, i.e., spin current
generated by the anomalous Hall is fully polarized such that the
spin current is same as the charge current. Since the charge current
is completely blocked by the MIL, it is inevitable that the spin
current is also being completely blocked.

\subsection{NM$\mid$MIL bilayers}

In this section, we consider a NM$\mid$MIL bilayer. In this case,
the magnon current in the MIL is induced by a temperature gradient,
see Fig.~2(c). From the magnon Boltzmann equation within the
relaxation time approximation, the non-equilibrium magnon
distribution is,
\begin{equation}
\delta N_{\mathbf{q}} = -v_{\mathbf{q}}^x \tau_m \frac{\partial N_{\mathbf{q}%
}^0}{\partial T} \frac{dT}{dx} - v_{\mathbf{q}}^x \tau_m \frac{\partial
\delta N_{\mathbf{q}} }{\partial x}.
\end{equation}
where $v_{\mathbf{q}}^x$ denotes the $x$-component of the magnon
velocity and $\tau_m$ is the magnon-conserving relaxation time.
By defining the magnon current as $j_m \equiv (2\mu_B)\int d\mathbf{q} v^{x}_{\mathbf{%
q}} \delta N_{\mathbf{q}}$ and following the derivation in the
Appendix A of the Supplemental Material of Ref.~\cite{Zhang12}, we
find
\begin{equation}
j_{m}=- \kappa\frac{dT}{dx}-D_{m}\frac{\partial \delta m_{m}}{\partial x}
\end{equation}%
with
\begin{equation}
\kappa = \frac{2\sqrt{3(S+1)} \mu_B \tau_m k_{B}^{2}T_c \xi}{9 \pi \hbar^2
a_{0I}} \left( \frac{T}{T_c} \right)^\frac{3}{2}
\end{equation}
and $\xi = \int^{\infty}_{0}dx x^{3/2}e^{x}/(e^x-1)^2\simeq 3.4$.
The magnon accumulation satisfies the magnon diffusion equation
whose solution can be taken as a simple form,
\begin{equation}
\delta m_m (x) = B_1 \exp(x/l_m)
\end{equation}
where we have assumed the thickness of the MIL to be much larger
than the magnon diffusion length $l_m$ and hence dropped the term
$\exp(-x/l_m)$ in the solution. The spin accumulation in the NM
layer can be written as
\begin{equation}
\delta m_s (x) = B_2 \exp(x/\lambda_{sf}) + B_3 \exp(-x/\lambda_{sf})
\end{equation}
and the spin current perpendicular to the plane is given by
$j_{s}=-D_{s}
\partial {\delta m_s}/{\partial x}$. These three integral constants
$B_i$ ($i=1,2,3$) can be determined by the two interface boundary
conditions Eqs.~(2) and (3), along with the outer boundary condition
$j_{s}(x=d+L_2)=0$. After a straightforward algebra, we get
\begin{equation}
j_{s}(x)=\frac{b \kappa \sinh \left( \frac{d+L_2-x}{\lambda
_{sf}}\right)}{(1+b
) \sinh \left( \frac{L_2}{\lambda _{sf}}\right) + a \cosh \left( \frac{L_2}{%
\lambda_{sf}} \right)} \frac{dT}{dx}
\end{equation}%
Again, the above perpendicular-to-plane spin current can generate an
in-plane electric current whose average density over the thickness
of the NM layer can be obtained by $\bar{j}_{e} =
(\gamma_{sh}/L_2)\int j_s(x) dx$. By taking the temperature gradient
as a constant, we have
\begin{equation}
\bar{j}_{e} = \frac{\lambda_{sf}}{L_2}\frac{\gamma_{sh} b \kappa
\left[ \cosh \left(
\frac{L_2}{\lambda_{sf}} \right) -1 \right]}{(1+b ) \sinh \left( \frac{L_2}{%
\lambda _{sf}}\right) + a \cosh \left( \frac{L_2}{\lambda_{sf}} \right)} \frac{%
d T}{dx}.
\end{equation}
The direction of the electric current is in the plane of the layer
and perpendicular to the directions of the magnetization as well as
the temperature gradient of the MIL. It is interesting to compare
the current driven electric drag, Eq.~(21), with the thermally
driven electrical drag, Eq.~(32). Firstly, in the former case, the
electric drag is proportional to the square of the Hall angle
because the first metallic layer converts the electric current to
the spin current via spin Hall effect and the second metal layer
converts the spin current into the electric current via inverse spin
Hall effect, while in the latter case, the spin current is directly
injected from the thermally driven magnon current and thus the drag
current is linearly proportional to the spin Hall
angle. Secondly, in the NM$\mid$MIL$\mid$NM case, both spin convertances $G_{em}$ and $%
G_{me}$ are important, while for NM$\mid$MIL, the convertance
relating the magnon accumulation to the spin current, $G_{me}$,
plays a dominant role. A rough estimation yields the induced current
density in a
Pt$\mid$YIG bilayer is about $10$ $A/cm^2$ for a moderately small temperature gradient of $dT/dx=10$ $%
K/cm$ if one chooses the following parameters: $%
\gamma_{sh}=0.05$ \cite{Buhrman11}, $J_{sd}=1$ $meV$,
$L_2=\lambda_{sf}=7$ $nm$, $c_{Pt}=0.1$ $(\mu \Omega\cdot cm)^{-1}$,
$a_{0M}=3.9$ ${\mathring{A}}$, $a_{0I}=12.376$ ${\mathring{A}}$,
$\tau_m=10^{-8} $ $s$ \cite{Callen65}, $\tau_{th}=10^{-6}$ $s$,
$l_m=1$ $mm$, $S=\frac{5}{2}$ and $T_c=550$ $K$.

\section{Discussions and Summary}

We have investigated the spin transport across the interface between
a metal layer and a MIL. The salient feature of our approach is that
we have treated spin and magnon transport properties on an equal
footing. Namely, the spin and magnon accumulations as well as the
spin and magnon currents are described by semiclassical
non-equilibrium distribution
functions. In other approaches, for example, Xiao \emph{et al.} \cite%
{Uchida10,Xiao10} described the magnon density through a
quasi-equilibrium effective magnon temperature which differs from
the lattice temperature. Adachi \emph{et al.}
\cite{Adachi11,Adachi12} considered the linear response theories and
their numerical solutions \cite{Ohe} on the spin Seeback effect
\cite{Uchida10-nat, Uchida10-APL} in ferromagnetic insulators. These
approaches also provide alternative physical insights on the roles
of magnons in non-equilibrium transport \cite{Slonczewski10}.

We thank S. Bender for pointing out an inconsistency in the
approximations used in the derivation of the spin convertance
$G_{em}$ in a previous version of the paper. This work is supported
by NSF-ECCS.


\begin{thebibliography}{99}
\bibitem{Fert} M. N. Baibich , J. M. Broto, A. Fert, F. Nguyen Van Dau, F.
Petroff, P. Eitenne, G. Creuzet, A. Friederich, and J. Chazelas, Phys. Rev.
Lett. \textbf{61}, 2472 (1988).

\bibitem{Grunberg} G. Binasch, P. Grunberg, F. Saurenbach, and W. Zinn,
Phys. Rev. B \textbf{39}, 4828 (1989).

\bibitem{Slonczewski} J. C. Slonczewski, J. Magn. Magn. Mater. \textbf{159},
L1 (1996).

\bibitem{Berger} L. Berger, Phys. Rev. B \textbf{54}, 9353 (1996).

\bibitem{Hirsch} J. E. Hirsch, Phys. Rev. Lett. \textbf{83} 1834 (1999).

\bibitem{Zhang00} S. Zhang, Phys. Rev. Lett. \textbf{85}, 393 (2000).

\bibitem{Kajiwara10} Y. Kajiwara \emph{et al.}, Nature \textbf{464}, 262
(2010).

\bibitem{Xiao12} J. Xiao and G. E. W. Bauer, Phys. Rev. Lett. \textbf{108},
217204 (2012).

\bibitem{Wu11} Z. Wang, Y. Sun, M. Wu, V. Tiberkevich, and A. Slavin, Phys.
Rev. Lett. \textbf{107}, 146602 (2011).

\bibitem{Nowak11} D. Hinzke and U. Nowak, Phys. Rev. Lett. \textbf{107},
027205 (2011).

\bibitem{Wang11} P. Yan, X. S. Wang, and X. R. Wang, Phys. Rev. Lett.
\textbf{107}, 177207 (2011)

\bibitem{Zhang12} S. S.-L. Zhang and S. Zhang, Phys. Rev. Lett. \textbf{109}%
, 096603 (2012).

\bibitem{Valet93} T. Valet and A. Fert, Phys. Rev. B \textbf{48}, 7099
(1993).

\bibitem{Zhang&Parkin97} S. Zhang, P. M. Levy, A. C. Marley, and S. S. P.
Parkin, Phys. Rev. Lett. \textbf{79}, 3744 (1997).

\bibitem{Takahashi10} S. Takahashi, E. Saitoh, and S. Maekawa, J. Phys.:
Conference Series \textbf{200}, 062030 (2010).

\bibitem{Tserkovnyak12} S. A. Bender, R. A. Duine, and Y. Tserkovnyak, Phys. Rev. Lett. \textbf{108}, 246601 (2012).

\bibitem{Saitoh06} E. Saitoh, M. Ueda, H. Miyajima, and G. Tatara, Appl.
Phys. Lett. \textbf{88}, 182509 (2006).

\bibitem{Buhrman12} L. Liu, \emph{et al.}, Science \textbf{336}, 555 (2012).

\bibitem{Callen65} C. W. Haas and H. B. Callen, in Magnetism, Vol. I, edited
by G. T. Rado and H. Suhl (Academic Press, New York, 1965).

\bibitem{Buhrman11}L. Q. Liu, T. Moriyama, D. C. Ralph, and R. A. Buhrman,
Phys. Rev. Lett. \textbf{106}, 036601 (2011); a lower spin Hall
angle for Pt was reported in Z. Feng et al., Phys. Rev. B
\textbf{85}, 214423 (2012).

\bibitem{Uchida10} K. Uchida et al., Nature Mater. \textbf{9}, 894 (2010).

\bibitem{Xiao10} J. Xiao, G. E. W. Bauer, K. Uchida, E. Saitoh, and S.
Maekawa, Phys. Rev. B \textbf{81}, 214418 (2010).

\bibitem{Adachi11} H. Adachi, J. Ohe, S. Takahashi, and S. Maekawa, Phys. Rev.
B \textbf{83}, 094410 (2011).

\bibitem{Adachi12} H. Adachi and S. Maekawa, arXiv:1209.0228v1.

\bibitem{Ohe} J. Ohe, H. Adachi, S. Takahashi, and S. Maekawa, Phys. Rev. B
\textbf{83}, 115118 (2011)

\bibitem{Uchida10-nat}K. Uchida et al., Nature Mater. \textbf{9}, 894 (2010).

\bibitem{Uchida10-APL}K. Uchida et al., Appl. Phys. Lett. \textbf{97}, 172505 (2010).

\bibitem{Slonczewski10} J. C. Slonczewski, Phys. Rev. B \textbf{82}, 054403
(2010).
\end{thebibliography}
\end{document}